\documentclass[english,aps,prc,twocolumn,superscriptaddress]{revtex4}
\usepackage[T1]{fontenc}
\usepackage[latin9]{inputenc}
\usepackage{graphicx}
\usepackage{amssymb}
\usepackage{esint}
\usepackage{epstopdf}

\makeatletter

\usepackage{babel}
\makeatother

\begin{document}

%\preprint{This line only printed with preprint option}

\newcommand{\lmk}{$<< \clubsuit$}
\newcommand{\rmk}{$\clubsuit >>$}
\newcommand{\hetrd}{$^3$He }

\title{Limits on possible new spin-spin interactions between neutrons from measurements of the Longitudinal Spin  Relaxation Rate of Polarized $^{3}$He Gas}

\author{Changbo  Fu}
\affiliation{Center for Exploration of Energy and Matter, Indiana University, Bloomington, IN 47408}

\author{W. M. Snow}
\email[Corresponding author: ]{wsnow@indiana.edu}
\affiliation{Center for Exploration of Energy and Matter, Indiana University, Bloomington, IN 47408}

\date{\today}

\begin{abstract}
New particles with masses in the sub-eV range have been predicted by various theories beyond the Standard Model. Some can induce new spin-spin interactions between fermions. Existing constraints on such interactions between nucleons with mesoscopic ranges (millimeters to nanometers) are quite poor.  Polarized $^{3}$He gas is an especially clean system to use to constrain these possible new spin-spin interactions because the spin-independent atomic potential between helium atoms is well-characterized experimentally. The small effects from binary atomic collisions in a polarized gas from magnetic dipole-dipole and other possible weak spin-spin interactions which lead to spin relaxation can be calculated perturbatively. We compare existing measurements of the longitudinal spin relaxation rate $\Gamma_{1}$ of polarized  $^3$He gas with theoretical calculations and set a $1\sigma$ upper bound on the pseudoscalar coupling strength $g_p$ for possible new neutron-neutron dipole-dipole interactions of $g_p^{(n)}g_p^{(n)}/4\pi \le 1.7\times 10^{-3}$ for distances larger than 100 nm. We also set new direct limits on possible gravitational torsion interactions between neutrons.

\end{abstract}


\maketitle

\section{Introduction}

New spin-dependent interactions of nature with very weak couplings to matter and ranges of \lq\lq mesoscopic\rq\rq\ scale (millimeters to nanometers) are poorly constrained by experiment and are attracting more theoretical attention\cite{Moody84, Raf90, Ros00,Jae10}. Particles which might transmit such interactions are starting to be referred to as WISPs (weakly-interacting sub-eV particles)~\cite{Jae10}. Symmetries broken at a high energy scale generically lead to weakly-coupled light particles with long-range interactions through Goldstone's theorem. Theoretical attempts to explain dark matter and dark energy can also produce new weakly-coupled long-range interactions. In both cases there are many examples in which the new interactions are spin-dependent. The fact that the dark energy density of order (1 meV)${^4}$ corresponds to a $~100$ $\mu$m  length scale by dimensional analysis also encourages searches for new phenomena around this scale~\cite{Ade03,Ade09}. Taken together, these developments suggest that it is reasonable to conduct experimental searches for possible new spin-dependent interactions
which act over mesoscopic distance scales.  

Many searches for new spin-dependent interactions have been motivated by the idea of axions ~\cite{Pec77,Wei78, Wilczek78, Ros00, Raf90, You96}, which can induce a $P$-odd and $T$-odd  interaction between polarized and unpolarized particles proportional to ${\vec{s}} \cdot {\vec{r}}$, where ${\vec{r}}$ is the distance between the particles and ${\vec{s}}$ is the spin of the polarized particle. Several other ideas can generate exotic spin-dependent interactions ~\cite{Hehl76, Shapiro02, Hammond02, Kostelec04, Arkani04, Arkani05, Georgi07, Kostelec08}. However the idea to search for new spin-dependent interactions can be considered within a more general theoretical context. Dobrescu and Mocioiu~\cite{Dob06} recently performed a general classification of interactions between nonrelativistic spin $1/2$ fermions assuming only rotational invariance. This analysis emphasized the rich variety of possibilities for new spin-dependent interactions. Of the 16 different terms in the elastic scattering amplitude uncovered in this analysis, 15 involve either one or both of the spins of the fermions. 

In this paper we will consider constraints on possible new spin-dependent,  velocity-independent forces between nucleons. For one boson exchange between two nonrelativistic spin $1/2$ fermions there are 9 types of potentials involving both spins. Six depend on the relative velocities of the particles and the remaining three ($V_{2}, V_{3}$, and $V_{11}$ in the notation of Dobrescu and Mocioiu) are velocity-independent:

\begin{eqnarray}\label{eq.potential.V2} V_{2}=\frac{\hbar c}{4\pi r} \vec{\sigma}_1 \cdot \vec{\sigma}_2\,e^{-r/\lambda}, \end{eqnarray} \begin{eqnarray}\label{eq.potential.V3} V_{3}&=&\frac{\hbar^3}{4\pi m^{2} r^{3}c} \{(\hat{\mathbf{\sigma}}_1\cdot\hat{\mathbf{\sigma}}_2) \left(1+\frac{r}{\lambda}\right) \nonumber \\ &&-3(\hat{\mathbf{\sigma}}_1\cdot\hat{\mathbf{r}}) (\hat{\mathbf{\sigma}}_2\cdot\hat{\mathbf{r}}) \left(1+\frac{r}{\lambda}+\frac{r^2}{3\lambda^2}\right) \}e^{-r/\lambda}, \end{eqnarray} and \begin{eqnarray}\label{eq.potential.V11} V_{11}=\frac{\hbar^2}{4\pi m r^{2}} (\hat{\sigma}_1 \times  \hat{\sigma}_2 )\cdot \hat{r}  \left(1+\frac{r}{\lambda}\right)e^{-r/\lambda}, \end{eqnarray} where $m$ is the mass,  ${\vec{s}_{i}}=\hbar\hat{\sigma}_{i}/2$ is the spin of the polarized particle, $\hbar$ is Planck constant, $\lambda$ is the interaction range, and $\hat{\mathbf{r}}={\mathbf{r}}/r$ is the unit vector between the particles.  

The existing constraints on new spin-spin interactions between nucleons at distance scales below 1 cm are generally rather poor. It is not hard to understand why: for shorter-range interactions both the number of particles that can be brought within the range of the interaction becomes smaller and smaller, and the required precision with which one can understand the large backgrounds from the electromagnetic fields that accompany any macroscopically polarized medium becomes more and more difficult to achieve. Several measurements~\cite{Wine91, Gle08, Vas09} constrain $V_{2}$ and $V_{3}$ at relatively large distances. The best constraints at atomic distance scales come from the work of Ramsey~\cite{Ram79} who used spectroscopy in molecular hydrogen to constrain $V_{2}$ and $V_{3}$ interactions between protons. Recently Kimball and coauthors~\cite{Kim10} have used measurements~\cite{Soboll72, Borel03} and calculations~\cite{Wal89, Tscherbul09} of cross sections for spin exchange collisions between $^{3}$He and Na atoms to constrain $V_{2}$, $V_{3}$, and $V_{8}$ between neutrons and protons. $V_{8}$ is a spin-dependent and velocity-dependent potential of the form

\begin{eqnarray}
\label{eq.potential.V8} 
V_{8}=\frac{\hbar c}{4\pi r} (\vec{\sigma}_1 \cdot \vec{v})( \vec{\sigma}_2 \cdot \vec{v})\,e^{-r/\lambda} 
\end{eqnarray} 

where $\vec{v}$ is the relative velocity of the particles (such a potential can also influence atomic spin exchange collisions). In atomic spin exchange collisions the spin-dependent part of the interaction is a small perturbation on the dominant spin-independent atom-atom potential, and theoretical calculations of the spin-exchange cross section can be performed with high accuracy given sufficiently precise data on atomic potentials. The theoretical calculations are simpler for systems involving light atoms, and experimental data on spin exchange cross sections exist under conditions dominated by fast binary atom-atom collisions which minimize possible contributions from three-body collisions and the spin-rotation interaction~\cite{Wal89, Walker97}. In combination with existing constraints from spectroscopy in molecular hydrogen~\cite{Ram79} mentioned above, these authors were also able to set indirect constraints for new interactions between neutrons.  

Measurements on ensembles of polarized $^{3}$He gas atoms have been used in several recent studies which constrain monopole-dipole interactions\cite{Ser09, Ig09, Pok10, Fu10, Pet10, Fu11}, which involve the spin of one of the two particles. In this paper we show that polarized $^{3}$He can also be used to improve existing constraints on possible new nucleon spin-dependent interactions involving the spins of both particles. The $^{3}$He nucleus is isolated enough from external influences by the inert closed electron shell that other weak interactions involving the spin of the nucleus can manifest themselves. The interactions between the $^{3}$He atoms in a gas at room temperature are dominated by binary atomic collisions whose dynamics have been accurately calculated using the well-measured He-He atomic potential, and the extra effects from weak spin-dependent interactions can be treated to high accuracy as weak perturbations. Unlike the spin exchange collisions between noble gas atoms and alkali metal atoms, there is no contribution from polarized electrons. Experimental measurement coupled with theoretical analysis shows that the polarization of the $^{3}$He nucleus is dominated as one would expect by the polarization of the neutron~\cite{Friar90}, and therefore any limit derived from this system can be attributed directly to a limit on neutron-neutron interactions. 

The spin exchange cross section between $^{3}$He atoms can be calculated with relatively high accuracy using the well-measured atomic potentials since (as for Na-$^3$He) the spin-dependent part of $^3$He-$^3$He scattering is also a small perturbation on the dominant spin-independent part. The spin relaxation rate $\Gamma_1^{(1)}$ is simply related to the $^3$He-$^3$He spin-exchange cross section $\sigma_{1,E}^{(1)}$\cite{Chapman75}

\begin{eqnarray}\label{eq.gamma.V1}
\Gamma_1^{(1)}=n
\left(
\frac{2}{\pi \mu (k_B T)^{3}}
\right)^{1/2}
\int_0^\infty e^{-E/k_B T}\sigma_{1,E}^{(1)}\,E{\rm d}E,
\end{eqnarray}
where $k_B$ is the Boltzmann constant, $E$ is the energy of the particles, $\mu$ is the reduced mass, $T$ is the temperature, and $n$ is the gas density. Furthermore, there is extensive data on the longitudinal spin relaxation rate $\Gamma_{1}$ of ensembles of polarized $^{3}$He gas atoms under conditions in which this rate is dominated by binary  $^{3}$He-$^{3}$He spin exchange collisions.  By using special glass cells to suppress the loss of polarization from interaction with the container walls, the measured spin relaxation rate of polarized $^{3}$He gas in certain cells is so slow (relaxation times on the order of several hundred hours) that the measured rate closely approaches the rate calculated from magnetic dipole-dipole interactions. Since the events which lead to the $\Gamma_{1}$ relaxation rate come from a large number of binary atom-atom collisions between many pairs of polarized atoms, $\Gamma_{1}$ measurements have the potential to be more sensitive to new interactions than measurements of spin exchange cross sections, which involve single binary collisions between atoms.

In this work, we compare the measured longitudinal spin relaxation rates $\Gamma_{1}$ of polarized $^3$He gas with theoretical calculations of  $\Gamma_{1}$ from magnetic dipole-dipole interactions to set a limit for possible new spin-spin couplings between neutrons. The rest of this paper is organized as follows. In Sec.~II we discuss the physical mechanisms which can lead to $\Gamma_{1}$ spin relaxation in an ensemble of polarized gas atoms and argue that the dominant contributions for the data considered in this paper come from spin exchange collisions and interactions of the polarized atoms with the cell walls. We also outline the calculation of the contribution to $\Gamma_{1}$ from the magnetic dipole-dipole interaction. The spin dependent potential $V_{3}$ described above is directly proportional to the magnetic dipole-dipole interaction in the $\lambda \to \infty$ limit. We observe that there is a distance scale beyond which the difference between the radial dependence of the matrix elements involved in the calculation of the spin exchange cross section for the two interactions is negligible. In Sec.~III we present experimental data on $\Gamma_{1}$ spin relaxation rates and use this data to set limits on the pseudoscalar coupling $g_{p}$ which generate the $V_{3}$ potential and on possible contributions from gravitational torsion between neutrons. Sec.~IV has our conclusions and suggestions for further work. 

\section{$\Gamma_{1}$ Spin Relaxation Mechanisms in Polarized $^{3}$He Gas Cells}

Interactions which can cause longitudinal spin relaxation in an ensemble of polarized $^{3}$He atoms include: (1) a possible electric dipole moment, (2) the spin-rotation interaction in $^{3}$He-$^{3}$He collisions, (3) wall relaxation($\Gamma_{1}^{(wall)}$), (4) magnetic field gradients($\Gamma_{1}^{(\partial B)}$),  (5) magnetic dipole-dipole interactions($\Gamma_{1}^{(1)}$), and (6) a new dipole-dipole interaction($\Gamma_{1}^{(2)}$). The experimental upper bounds on atomic electric dipole moments in general and on $^{3}$He in particular make mechanism (1) utterly negligible~\cite{Purcell60} and we shall not consider it further.

The spin-rotation interaction (mechanism 2 above) is proportional to ${\vec S} \cdot {\vec N}$ with ${\vec N}$ the orbital angular momentum coming from the motional magnetic fields seen by the polarized nucleus during atomic collisions~\cite{Walker97}. The interatomic interaction distorts the charge distribution of the atom and creates a fluctuating field. At room temperature the average kinetic energy of the colliding $^3$He atoms is small compared to the atomic binding energy and therefore these distortions are relatively small. In addition, the $^{3}$He-$^{3}$He interaction is weak enough that there are no molecular bound states which can allow the atoms to experience several revolutions and amplify the spin-rotation interaction, as happens for other atomic species. This effect has been investigated~\cite{Chapman75} for $^{3}$He and is negligible (<1\%) compared with magnetic dipole-dipole interaction at room temperature. 

The longitudinal relaxation rate $\Gamma_{1}^{(wall)}$ due to atomic collisions with the cell wall is significant. The detailed physics involved in the wall relaxation remains poorly understood. Nevertheless careful preparation of the surfaces of certain types of aluminosilicate glasses can reduce the relaxation rate from this process to be small compared to dipole-dipole relaxation, as will be shown in section~III.
 
Spin relaxation $\Gamma_{1}^{(\partial B)}$ from the motion of the polarized nuclei in magnetic field gradients can be calculated to be\cite{Cat88}
\begin{equation}
 \Gamma_1^{(\partial B)}=D\frac{|\triangledown B_{\bot}|^2}{B_{||}^2},
\end{equation}
where $D$ is the diffusion constant of the polarized gas and $B_{||}\ (B_{\bot})$ are the magnetic fields parallel (perpendicular) to the spin's direction. For a polarized $^3$He cell of pressure 1 bar at room temperature in an external field gradient of ${\delta B_{\bot}/ B_{||}}=10^{-4}$ cm$^{-1}$, which is  typically achieved in the Helmholtz coil arrangements used in the measurements described in section III,  $\Gamma_{1}^{(\partial B)}=2\times10^{-8}$~s$^{-1}$\cite{McIver09}, which is small compared to the dipole-dipole relaxation in the cells discussed in section III. Internal field gradients induced by the magnetization of the polarized gas are proportional to the gas density and polarization and also depend on the geometry of the gas container.  A field generated by polarized gas in a spherical cell is uniform and thus has no effect on spin relaxation. For the gas cells discussed in this paper the relaxation from self-generated internal field gradients is very small compared to relaxation from external field gradients.

The spin relaxation $\Gamma_{1}^{(1)}$ due to magnetic dipole-dipole interactions in binary $^{3}$He-$^{3}$He collisions dominates the bulk relaxation in the gas.  It can be calculated by first solving for the $^3$He-$^3$He scattering amplitude using the measured spin-independent atom-atom potential $V^{(0)}$ and then adding the hyperfine interaction as a perturbation.  The spin-dependent magnetic dipole-dipole interaction potential $V^{(1)}$ has the form
\begin{eqnarray}\label{eq.dipoledipole}
V^{(1)}(r)= 
\frac{f^{(1)}}{r^3}\left[
(\hat{\mathbf{\sigma}}_1\cdot\hat{\mathbf{\sigma}}_2)
-
3(\hat{\mathbf{\sigma}}_1\cdot\hat{\mathbf{r}})
(\hat{\mathbf{\sigma}}_2\cdot\hat{\mathbf{r}})
\right],
\end{eqnarray}
where $f^{(1)}=\alpha\frac{\hbar^3g^2}{16m_e^2c}$, $g=-0.002317$ is g-factor of  $^3$He, $\alpha$ is the fine structure constant, and $m_e$ is the mass of electron. The expression for $\Gamma_{1}$ from binary collisions of polarized atoms in a gas in thermal equilibrium at temperature $T$ is shown in Eq.~\ref{eq.gamma.V1}, where the spin exchange cross section can be written as
\cite{Mullin90,Shizgal73,New93},
\begin{eqnarray}\label{eq.gamma1k.normal}
\sigma_{1,E}^{(1)}&=&\frac{48\pi m^2}{5\hbar^4}  
\left( f^{(1)} \right)^2 \nonumber \\
&&\times \sum_{ll' (odd)}(2l+1)(2l'+1)C^2(ll'2;00)
\left<  \frac{1}{(kr)^3}  \right>^2_{ll'}.
\end{eqnarray}
where $C(ll'2;00)$ are Clebsch-Cordon coefficients, $k$ is the wavenumber,  and the matrix elements $\left< \frac{1}{(kr)^3}\right>_{ll'}$ corresponding to  $l\rightarrow l'$ transitions can be obtained by solving the Schrodinger equation for the two-body scattering states\cite{New93}.  

For atom-atom collisions at room temperature only partial waves with small~$l$ make significant contributions to the spin exchange cross section. Equating the centrifugal barrier at a distance corresponding to the $^{3}$He atomic diameter of $0.06$ nm with the kinetic energy, $l(l+1)\hbar^2/2\mu r^2=3k_B T/2$ (where $\mu$ is the reduced mass of $^3$He) to find the orbital angular momentum associated with the closest approach of the atoms yields $l\simeq3$. At room temperature partial waves with $l>3$ do not penetrate the centrifugal barrier and therefore see mainly the long-range Van der Waals interaction.

The potential energy from the possible new dipole-dipole interaction which we propose to constrain has the form\cite{Moody84},
\begin{eqnarray}\label{eq.potential.org}
V^{(2)} (r)&=&
f^{(2)}
\frac{e^{-r/\lambda}}{r^3}
\{(\hat{\mathbf{\sigma}}_1\cdot\hat{\mathbf{\sigma}}_2)
\left(
1+\frac{r}{\lambda}
\right) 
\nonumber \\
&&-3(\hat{\mathbf{\sigma}}_1\cdot\hat{\mathbf{r}})
(\hat{\mathbf{\sigma}}_2\cdot\hat{\mathbf{r}})
\left(
1+\frac{r}{\lambda}+\frac{r^2}{3\lambda^2}
\right)
\},
\end{eqnarray}
where $f^{(2)}=g_{p}^2\hbar^{3}/(16\pi m_n^2c)$, $m_n$ is the mass, $c$ is the speed of light, $\lambda$ is the interaction range, and $g_{p}$ is the coupling constant. Because $V^{(2)}< V^{(1)}\ll V^{(0)}$, $V^{(2)}$ can also be treated as a perturbation and one can follow the same procedure as in Eq. (\ref{eq.gamma1k.normal}) to obtain the matrix elements of $\left< V^{(2)}\right>_{ll'}$. Note that the dipole-dipole potential under exchange of a finite-mass particle in Eq.~(\ref{eq.potential.org}) reduces to the same form as the usual electromagnetic dipole-dipole potential in Eq.~(\ref{eq.dipoledipole}) as the particle becomes massless ($\lambda \to \infty$). In this limit the analysis required to set a bound on $V_{3}$ is greatly simplified. For the data considered in this paper this limiting case is reached already for interaction ranges $\lambda>100$ nm. The interatomic potential $V^{(0)}(r) $ falls quickly outside the atomic diameter, and collisions with impact parameters of this size between $^{3}$He atoms at room temperature correspond to $l \approx 3$. Partial waves with $l>3$ feel only the weak long-range part of the atom-atom potential, which can be calculated in perturbation theory and makes a small contribution to the matrix element. The lower partial waves encounter the hard core repulsion. For interaction ranges $\lambda>100$ nm, however, the Yukawa term in the potential is slowly varying and the radial dependence of $V^{(2)}$ and $V^{(1)}$ are therefore the same to high accuracy for small $r$. One can show numerically that beyond $r=10$ nm the lower partial waves approach their asymptotic forms and make negligible contributions to the matrix element. Therefore one can choose an upper cutoff of $r=10$ nm in the radial integral for the calculation of matrix element $\left< V^{(2)}\right>_{ll'}$ with negligible uncertainty.  In this case $\lambda\gg r$ in the matrix elements  and Eq.~(\ref{eq.potential.org}) can be simplified to
\begin{eqnarray}\label{eq.potential.simplified}
V^{(2)}(r)&\simeq&
\frac{f^{(2)}}{r^3}\left[
(\hat{\mathbf{\sigma}}_1\cdot\hat{\mathbf{\sigma}}_2)
-
3(\hat{\mathbf{\sigma}}_1\cdot\hat{\mathbf{r}})
(\hat{\mathbf{\sigma}}_2\cdot\hat{\mathbf{r}})
\right].
\end{eqnarray}
In this limit $V^{(2)}$ and $V^{(1)}$ have the same form, so the contribution of the potentials $V^{(1)}$ and $V^{(2)}$ to the longitudinal spin relaxation rate is
\begin{eqnarray}\label{eq.gamma.V12}
\Gamma_1^{(1,2)}=n
\left(
\frac{2}{\pi \mu (\kappa T)^{3}}
\right)^{1/2}
\int_0^\infty e^{-E/\kappa T}\sigma_{1,E}^{(1,2)}\,EdE,
\end{eqnarray}
where $\sigma_{1,E}^{(1,2)}$ is
\begin{eqnarray}\label{eq.gamma12k.normal}
\sigma_{1,E}^{(1,2)}&=&\frac{48\pi m^2}{5\hbar^4}  
\left( f^{(1)}+f^{(2)} \right)^2 \nonumber \\
&&\times \sum_{ll' (odd)}(2l+1)(2l'+1)C^2(ll'2;00)
\left<  \frac{1}{(kr)^3}  \right>^2_{ll'}.
\end{eqnarray}
Using Eq.~(\ref{eq.gamma.V1}) and 
Eq.~(\ref{eq.gamma.V12}), we have,
\begin{equation}\label{eq.gp.0}
 g_{p}^2/4\pi
=\frac{\alpha\,g^2(m_n/m_e)^2 }{4} \left[\left(\frac{\Gamma_1^{(1,2)}}{\Gamma_1^{(1)}}\right)^{1/2}-1\right].
\end{equation}

The experimentally measured longitudinal relaxation rate $\Gamma_1^{(exp)}$ must satisfy $\Gamma_1^{(exp)}\ge\Gamma_1^{(1,2)}$. Therefore, using Eq. (\ref{eq.gp.0}) we can derive a lower limit for the product of the couplings $g_{p}^{2}$: 
 
\begin{eqnarray}\label{eq.gp.exp}
 g_{p}^2/4\pi
&\le&\frac{\alpha\,g^2(m_n/m_e)^2 }{4} \left[\left(\frac{\Gamma_1^{(exp)}}{\Gamma_1^{(1)}}\right)^{1/2}-1\right] \nonumber \\
&= & 0.033 R,
\end{eqnarray}
where $R\equiv \left[\left(\Gamma_1^{(exp)}/\Gamma_1^{(1)}\right)^{1/2}-1\right] $ is an upper bound on the strength of the new dipole-dipole interaction relative to the magnetic dipole-dipole interaction: $V^{(2)}/V^{(1)}\le R$. The uncertainty on the constraint on $g_{p}^2/4\pi$ is determined by the experimental uncertainty of $\Gamma_1^{(exp)}$ and the theoretical uncertainty in the calculated value for $\Gamma_1^{(1)}$. 

\section{Measurements of $\Gamma_{1}^{(exp)}$ Relaxation Rates of Polarized $^3$He Gas}

The technology of laser optical pumping to produce macroscopic quantities of gas with high polarization has undergone extensive development for scientific applications in neutron scattering, medical imaging, and nuclear and particle physics~\cite{Babcock09, Chen07, Babcock06, Holmes08, Slifer08}. There exist two widely-used methods to polarize $^3$He gas, metastability-exchange optical pumping (MEOP) \cite{Colegrove63} and spin-exchange optical pumping (SEOP) \cite{Walker97}. We use SEOP data in this paper. In addition to the $^{3}$He, SEOP cells also contain a small amount (normally <0.1 g) of Rb and/or K for optical pumping and a small amount of N$_2$ gas to nonradiatively relax the optically-pumped alkali atoms to prevent radiation trapping. It has been found experimentally that certain aluminosilicate glasses can have wall relaxation rates which are small compared to the dipole-dipole relaxation rate, and it is $\Gamma_{1}^{(exp)}$ measurements in these cells that we use to set our limits. 

To our knowledge the most accurate comparisons between theory and experiment for the $\Gamma_{1}$ spin relaxation rates of polarized $^3$He gas in a SEOP cell were performed by Newbury et al.~\cite{New93, Newthesis} and Rich et al. ~\cite{Rich02}. For $^3$He at room temperature $\Gamma_1^{(1)}$ is~\cite{New93}
\begin{equation}
 \Gamma_1^{(1)}=3.73\times 10^{-7}\cdot [n] \cdot {\rm s^{-1}},
\end{equation}
where the $^3$He density $[n]$ is in units of amagats. Some of the theoretical uncertainties in this result come from the uncertainty in the measurements of the interatomic potential (which differ slightly\cite{Aziz87}) and produce corresponding uncertainties in the calculated spin exchange rate. The relaxation rates calculated with these different interatomic potentials $V^{(0)}$ differ by 1-2\%. We assign a 2\% relative standard uncertainty in $\Gamma_{1}^{(1)}$ from experimental knowledge in interatomic potentials and a 1\% relative standard uncertainty from other sources related to the numerical calculation. In addition there are uncertainties in the theoretical prediction associated with uncertainties in the knowledge of the temperature and density of the $^{3}$He in the cells.  In Ref. \cite{New93}, the densities of the cells are 8.37(19) amagat for cell {\it 808}, and 4.67(11) amagat for cell {\it 842}, where the numbers in parentheses denote the standard uncertainties in the last digit(s).  $\Gamma_{1}^{(1)}$ is proportional to the density, and near room temperature $\Gamma_1\propto T^{1/2}$\cite{Mullin90}.
We assume a temperature and corresponding standard uncertainty of $(297\pm5)\ ^\circ$C,  which yields a $\Gamma_1^{(1)}$ relative standard uncertainty of 0.8\%. The relative standard uncertainty in the theoretical prediction for $\Gamma_{1}$ for the Newbury cells was therefore ${\Delta\Gamma_{1}^{(1)} \over \Gamma_{1}^{(1)}}=3\%$.  The measured relaxation rates are $3.19(4)\times10^{-6}$/s for cell {\it 808} and $1.84(2)\times10^{-6}$/s for cell {\it 842}. Using these numbers and equation \ref{eq.gp.exp}, we set a limit of $g_p^{(n)}g_p^{(n)}/4\pi<1.7 \times10^{-3}$ at the $1\sigma$ confidence level. 

A similar upper bound is also obtained in an independent measurement at a lower gas density. In Ref. \cite{Rich02}, the cell Wilma with gas density of  0.73(3) amagat (the original paper mistakenly stated 0.78 amagat for the density) was measured to have spin relaxation rate $3.31(6)\times10^{-7}$/s. Using these numbers in equation \ref{eq.gp.exp} would set a slightly less stringent limit of $g_p^{(n)}g_p^{(n)}/4\pi<5\times10^{-3}$ at the $1\sigma$ confidence level. 
We also note that $\Gamma_{1}^{(exp)}$ measurements for several other $^{3}$He SEOP cells have been conducted over the last decade~\cite{Smith98, JCNS2010} with densities ranging from $0.7$ amagat to $2$ amagat and with different types of glasses as the cell materials.  Although many of these cells possess wall relaxation rates comparable to those discussed above, none of the $\Gamma_{1}^{(exp)}$ measurements in these cells is less than the calculated $\Gamma_1^{(1)}$. 

\begin{figure}
\begin{center}
\includegraphics[width=8.cm]{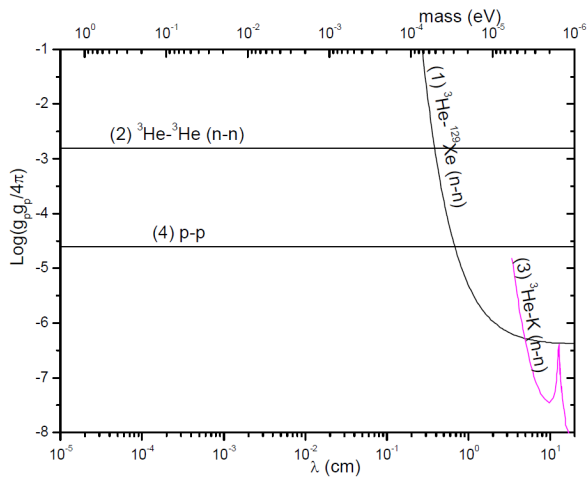}
\caption{\label{fig:result} Comparison of $1\sigma$ upper bounds on the coupling constant combination $g_p^{2}/4\pi$ for possible new pseudoscalar dipole-dipole interactions from the following sources: 1).  using $\Gamma_{1}$ measurements in a $^3$He-$^{129}$Xe maser\cite{Gle08}, 2). this work, using measurements in $^{3}$He SEOP cells,   3). using interaction between $^3$He and K species in two separate SEOP cells\cite{Vas09}, and 4). using hydrogen molecular spectroscopy\cite{Ram79}.
} 
\end{center}
\end{figure}

In Figure 1 we show the limits on $g_{p}^2/4\pi$ for neutrons extracted via Eq.~(\ref{eq.gp.exp}) using the comparison between the $\Gamma_{1}^{(exp)}$ measurements with theory. Limits from measurements in a $^3$He-$^{129}$Xe maser\cite{Gle08} and from measurements involving separate SEOP cells of $^3$He and K\cite{Vas09} are the most stringent for neutron-neutron interactions at distances greater than 1 cm. Hydrogen molecular spectroscopy\cite{Ram79} provides a stringent direct constraint on proton-proton interactions. The indirect limits on neutron-neutron interactions set by Kimball et al. on $g_{p}^2/4\pi$ for the $V_{3}$ interaction are above the top of the vertical axis of the plot. The limits from $^{3}$He-$^{3}$He, which as mentioned in the introduction are cleanly interpretable as direct limits on neutron-neutron interactions, are the best direct limits to our knowledge for distance scales from 100 nm to a few mm, corresponding to exchange particles with masses from 1 eV to 0.1 meV. 

We can also use this data to constrain short-range gravitational torsion between neutrons. Torsion, an additional warping of spacetime with spin as its source, is required for the conservation of angular momentum in general relativity when intrinsic spin is included~\cite{Hammond10}. Recently the experimental constraints on long-range gravitational torsion have been tightened considerably with the realization~\cite{Kostelec04} that a background torsion field violates effective local Lorentz invariance. Kosteleck\'y and coauthors ~\cite{Kostelec08} were able to constrain $19$ of the $24$ components of the torsion tensor for the first time, and these methods have since been adopted~\cite{Heckel08} to further improve constraints on $4$ of these $19$ components. As for short-range spin-dependent interactions, experimental constraints on short-range torsion~\cite{Neville80, Neville82, Carroll94, Hammond95} are poor. The spin-spin interaction generated by a short-range torsion field is of the same form ($V_{3}$ in the Dobrescu notation) considered above~\cite{Hammond95, Ade09} scaled by a parameter $\beta$ where

\begin{eqnarray}\label{torsion}
\beta^{2}=(g_{p}^2/4\pi \hbar c)({{2 \hbar c} \over 9Gm_{n}^{2}})
\end{eqnarray}

Figure 2 shows the constraints on short-range torsion from the work of Kimball et al~\cite{Kim10},  Ramsey~\cite{Ram79}, and our work, which limits $\beta^{2} < 2 \times 10^{37}$ for neutron-neutron interactions. Over the distance range between 100 nm and 1 cm these works set to our knowledge the best experimental limits on the torsion parameter for neutron-proton, neutron-neutron, and proton-proton interactions, respectively. At distance scales $> 50$ cm a much more stringent limit of $\beta^{2} < 2 \times 10^{28}$ comes from the work of  Romalis and co-workers~\cite{Vas09}.

\begin{figure}
\begin{center}
\includegraphics[width=8.cm]{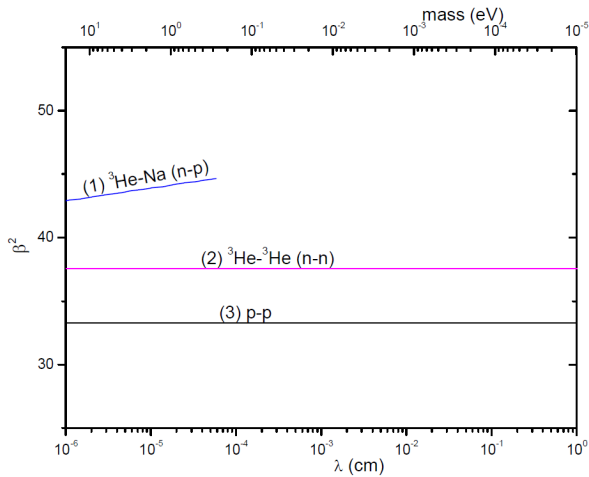}
\caption{\label{fig:result} Comparison of $1\sigma$ upper bounds on gravitation torsion couplings between nucleons from the following sources: 1).  from the analysis of  $^3$He-Na spin exchange cross section measurements\cite{Kim10}, 2). this work,  3). using hydrogen molecular spectroscopy\cite{Ram79}.
} 
\end{center}
\end{figure}

\section{Conclusions}

By comparing  theory and experiment for the longitudinal spin relaxation rate $\Gamma_{1}$ of polarized $^{3}$He gas, we have derived a $1\sigma$ upper bound of $g_p^{(n)}g_p^{(n)}/4\pi < 1.7\times 10^{-3}$ on the product of the couplings for a possible new pseudoscalar spin-spin interaction between neutrons with a dipole-dipole form $V_{3}$ at distance scales larger than 100 nm. This constraint limits such interactions to a size no more than 4\% of the magnetic dipole-dipole interactions between the polarized $^{3}$He atoms. Although these limits are certainly much less stringent than those for spin-spin interactions of macroscopic range, they are to our knowledge the best direct limits from laboratory experiments on spin-spin interactions of this form between neutrons. Attempts to improve the constraints using this approach would be limited both by the spin exchange cross section uncertainties and our ignorance of the physics of $^{3}$He wall relaxation rates, and we do not expect that it will be possible to significantly improve the accuracy of either the theoretical calculations or the understanding of the physics of the wall relaxation in the near future. However it should be possible to use this same data to set the best constraints on the $V_{2}$ and $V_{11}$ spin-dependent potentials and the velocity and spin-dependent potential $V_{8}$  for distance scales larger than 100 nm. We plan to present these limits in a forthcoming publication.

 \section{Acknowledgements}

We thank Tom Gentile, Wangchun Chen, Xilin Zhang for extensive discussions. This work was supported in part by NSF PHY-0116146. The work of WMS was supported in part by the Indiana University Center for Spacetime Symmetries.

%\bibliography{5forceCitations}

\begin{thebibliography}{}
\bibitem{Raf90} G. G. Raffelt, Physics Reports {\bf 198}, 1 (1990). 
\bibitem{Ros00} L. J. Rosenberg and K. A. van Bibber, Physics Reports 
{\bf 325}, 1 (2000). 
\bibitem{Moody84} J. E. Moody and F. Wilczek, Phys. Rev. D {\bf 30}, 130 (1984). 
\bibitem{Jae10}  J. Jaeckel and A. Ringwald, Annu. Rev. Nucl. Part. Sci. {\bf 60}: 405 (2010). 
\bibitem{Ade03} E. G. Adelberger, B. R. Heckel, and A. E. Nelson, Annu. 
Rev. Nucl. Part. Sci. {\bf 53}, 77 (2003). 
\bibitem{Ade09}  E. G. Adelberger {\it et al.}, Prog. Part. Nucl. Phys. {\bf 62}, 102 
(2009). 
\bibitem{Pec77} R. D. Peccei and H. R. Quinn, Phys. Rev. Lett. {\bf 38}, 1440 
(1977). 
\bibitem{Wei78} S. Weinberg, Phys. Rev. Lett. {\bf 40}, 223 (1978).
\bibitem{Wilczek78} F. Wilczek Phys. Rev. Lett. {\bf 40}, 279 (1978). 
\bibitem{You96} A. N. Youdin, D. Krause, K. Jagannathan, L. R. Hunter, S. K. Lamoreaux,  Phys. Rev. Lett. {\bf 77}, 2170 (1996).
\bibitem{Hehl76} F. W. Hehl, P. von der Heyde, G. D. Kerlick, and J. M. Nester, Rev. Mod. Phys. {\bf 48}, 393 (1976).
\bibitem{Shapiro02} I. L. Shapiro, Phys. Rep. {\bf 357}, 113 (2002).
\bibitem{Hammond02} R. T. Hammond, Rep. Prog. Phys. {\bf 65}, 599 (2002).
\bibitem{Kostelec04} V.A.\ Kosteleck\'y, Phys. Rev. D {\bf 69}, 105009 (2004).
\bibitem{Arkani04} N. Arkani-Hamed {\it et al.}, J. High Energy Phys. {\bf 05}, 074 (2004).
\bibitem{Arkani05} N. Arkani-Hamed, H. -C. Cheng, M. Luty, and J. Thaler, J. High Energy Phys. 07, 029 (2005).
\bibitem{Georgi07} H. Georgi, Phys. Rev. Lett. {\bf 98}, 221601 (2007).  
\bibitem{Kostelec08} V. A.\ Kosteleck\'y. N. Russell, and J. D. Tasson, Phys. Rev. Lett. {\bf 100}, 111102 (2008). 
\bibitem{Dob06} B. Dobrescu and I. Mocioiu, J. High Energy Phys. 0611, 
005 (2006). 
%\bibitem{Ni99} W.-T. Ni, S.-S. Pan, H.-C. Yeh, L.-S. Hou, and J. Wan, Phys. Rev. Lett. {\bf 82}, 2439 (1999). 
%\bibitem{Bae07} S. Baessler, V. V. Nesvizhevsky, K. V. Protasov, and A. Y. Voronin, Phys. Rev. D {\bf 75}, 075006 (2007). 
\bibitem{Wine91} D. J. Wineland, J. J. Bollinger, D. J. Heinzen, W. M. Itano, and M. G. Raizen, Phys. Rev. Lett. {\bf 67}, 1735 (1991).
\bibitem{Gle08} A. G. Glenday, C. E. Cramer, D. F. Phillips, and R. L. Walsworth, Phys. Rev. Lett. {\bf 101}, 261801 
(2008). 
\bibitem{Vas09} G. Vasilakis, J. M. Brown, T. W. Kornack, and M. V. Romalis,  Phys. Rev. Lett. {\bf 103}, 261801 (2009). 
\bibitem{Ram79} N. F. Ramsey, Physica A {\bf 96}, 285 (1979).
\bibitem{Kim10} D. F. Jackson Kimball, A. Boyd, and D. Budker, Phys. Rev. A {\bf 82}, 062714 (2010).
\bibitem{Soboll72} H. Soboll, Phys. Lett A {\bf 41}, 373 (1972).
\bibitem{Borel03} P. I. Borel, L. V. Sogaard, W. E. Svendsen, and N. Andersen, Phys. Rev. A {\bf 67}, 062705 (2003).
\bibitem{Wal89} T. G. Walker, Phys. Rev. A {\bf 40}, 4959 (1989).
\bibitem{Tscherbul09} T. V. Tscherbul, P. Zhang, H. R. Sadeghpour, and A. Dalgarno, Phys. Rev. A {\bf 79}, 062707 (2009).
\bibitem{Walker97} T. G. Walker and W. Happer, Rev. Mod. Phys. {\bf 69}, 629 (1997).
\bibitem{Ser09} A. Serebrov, Physics Letters B {\bf 680}, 423 (2009). 
\bibitem{Ig09} V. K. Ignatovich and Y. N. Pokotilovski, Eur. Phys. J. C 
{\bf 64}, 19 (2009). 
\bibitem{Fu10} C. Fu, T. R. Gentile, and W. M. Snow, Proceedings of 
the Fifth Meeting on CPT and Lorentz Symmetry, Bloomington, Indiana, 
June 28, 2010, ed. A. Kostelecky, p. 244 (2011); http://arxiv.org/abs/1007.5008.
\bibitem{Pok10} Y. N. Pokotilovski, Phys. Lett. B {\bf 686}, 114 (2010). 
\bibitem{Fu11} C. Fu, T. R. Gentile, and W. M. Snow, Phys. Rev. D83, 031504(R) (2011). 
\bibitem{Pet10} A. K. Petukhov, G. Pignol, D. Jullien, and K. H. Andersen, Phys. Rev. Lett. {\bf 105}, 170401 (2010). 
\bibitem{Friar90} J. L. Friar, B. F. Gibson, G. L. Payne, A. M. Bernstein, and T. E. Chupp,   Phys. Rev. C {\bf 42}, 2310 (1990).
\bibitem{Chapman75} R. Chapman, Phys. Rev. A {\bf 12}, 2333 (1975).
\bibitem{Purcell60} E. M. Purcell, Phys. Rev. {\bf 117}, 828 (1960). 
\bibitem{Cat88} G. D. Cates, S. R. Schaefer, and W. Happer, Phys. Rev. 
A {\bf 37}, 2877 (1988). 
\bibitem{McIver09} J. W. McIver {\it et al.}, Rev. Sci. Instrum. {\bf 86}, 063905(2009). 
\bibitem{Mullin90} W. J. Mullin,  F. Laloe, and M. G. Richards, J. Low Temp. Phys. {\bf 80}, 1 (1990).
\bibitem{Shizgal73} B. Shizgal, J. Chem. Phys. {\bf 58}, 3424 (1973).
\bibitem{New93} N. R. Newbury, A. S. Barton, G. D. Cates, W. Happer, and H. Middleton,  Phys. Rev. A. {\bf 48}, 4411 (1993).
\bibitem{Babcock09} E. Babcock {\it et al.}, Physica B {\bf 404}, 2655 (2009).
\bibitem{Chen07} W. C. Chen, T. R. Gentile, T. G. Walker, and E. Babcock, Phys. Rev. A {\bf 75}, 013416 (2007).
\bibitem{Babcock06} E. Babcock, B. Chaan, T. G. Walker, W. C. Chen, and T. R. Gentile, Phys. Rev. Lett. {\bf 96}, 083003 (2006).
\bibitem{Holmes08}  J. H. Holmes {\it et al.},  Magn. Reson. Med. {\bf 59}, 1062 (2008).
\bibitem{Slifer08} K. Slifer {\it et al.} (Jefferson Lab E94010 Collaboration), Phys. Rev. Lett. {\bf 101}, 022303 (2008).
\bibitem{Colegrove63} F.D. Colegrove, L.D.Schearer, and G.K. Walters, Phys. Rev. {\bf 132}, 2561(1963).
\bibitem{Newthesis} N. R. Newbury, PhD thesis, Princeton University (1992).  
\bibitem{Rich02}D. R. Rich {\it et al.}, App. Phys. Lett. {\bf 80}, 2210 (2002). 
\bibitem{Aziz87} R. A. Aziz,  F. R. W. McCourt, and C. C. K. Wong, Mol. Phys. {\bf 61}, 1487 (1987).
\bibitem{Smith98} T. B. Smith, T. E. Chupp, K. P. Coulter, and R. C. Welsh, Nucl. Inst. Meth. A {\bf 402}, 246 (1998).
\bibitem{JCNS2010} W.C. Chen, T.R. Gentile, C.B. Fu, S. Watson, G.L. Jones, J.W. McIver, and D.R. Rich, to be published in J. Phys. Conf. Series,  in press.
\bibitem{Hammond10} R. T. Hammond, Gen. Relativ. Grivit. {\bf 42}, 2345 (2010).
\bibitem{Heckel08} B. R. Heckel, E. G. Adelberger, C. E. Cramer, T, S. Cook, S. Schlamminger, and U. Schmidt,   Phys. Rev. D {\bf 78}, 092006 (2008).
\bibitem{Neville80} D. E. Neville, Phys. Rev. D {\bf21}, 2075 (1980).
\bibitem{Neville82} D. E. Neville, Phys. Rev. D {\bf 25}, 573 (1982).
\bibitem{Carroll94} S. M. Carroll and G. B. Field, Phys. Rev. D {\bf 52}, 6918 (1995).
\bibitem{Hammond95} R. T. Hammond, Phys. Rev. D {\bf 52}, 6918 (1995).
%\bibitem{McG90} D. D. McGregor, Phys. Rev. A {\bf 41}, 2631 (1990).
%\bibitem {PIS55}  D. Pines and C.P. Slichter, Phys. Rev. {\bf 100}, 1014 (1955).  
%\bibitem{Gen01} T. R. Gentile et al., J. Res. Natl. Inst. Stand. Technol. {\bf 106}, 709 (2001). 
%\bibitem{Ote92}  E. R. Oteiza, Ph.D. thesis, Harvard University, Cambridge, Massachusets (1992). 
%\bibitem{Chu88} T. E. Chupp, E. R. Oteiza, J. M. Richardson, and T. R. White, Phys. Rev. A {\bf 38}, 3998 (1988). 
%\bibitem{Bl46} F. Bloch, Phys. Rev. {\bf 70}, 460 (1946)
%\bibitem {BLO57}  S. Bloom, J. Appl. Phys. {\bf 28}, 800 (1957). 
%\bibitem{Rom01} M. V. Romalis and M. P. Ledbetter, Phys. Rev. Lett. {\bf 87}, 067601 (2001).
%\bibitem{Wal10} T. G. Walker, I. A. Nelson, and S. Kadleck, Phys. Rev. A {\bf 81}, 032709 (2010).
%\bibitem{Ham07} G. D. Hammond, C. C. Speake, C. Trenkel, and A. P. Paton, Phys. Rev. Lett. {\bf 98}, 081101 (2007). 
%\bibitem{Rit93} R. C. Ritter, L. I. Winkler, and G. T. Gillies, Phys. Rev. Lett. {\bf 70}, 701 (1993). 
%\bibitem{Rom98} M. V. Romalis and G. D. Cates, Phys. Rev. A {\bf 58}, 3004 (1998).
\end{thebibliography}
%\bibliographystyle{apsrev}

\end{document}